\documentclass{appolb}%
\usepackage{epsfig}
\usepackage{amsmath}

\newcommand{\w}{\vrule height 13 pt depth 0 pt width 0 pt}

\begin{document}

\title{Successes and Problems of Chiral Soliton Approach 
    to Exotic Baryons}%
\title{Successes and problems of chiral soliton approach \\
      to exotic baryons}
\author{Micha\l\ Prasza\l owicz$^a$ and Klaus Goeke$^b$
\address{
$^a$~M. Smoluchowski Institute of Physics,
 Jagellonian University,\\
Reymonta 4, 30-059 Krak{\'o}w, Poland\\
$^b$~Institut f\"ur Theoretische Physik II, Ruhr-Universit\" at Bochum,\\
         D--44780 Bochum, Germany
}} 
\maketitle

\begin{abstract}
We briefly review the formulation of chiral quark soliton model
and explain the difference and similarities with the Skyrme model.
Next, we apply the model to calculate non-exotic and exotic mass
spectra. We concentrate on large $N_{c}$ counting both for mass
splittings and decay widths. It is shown that pure large $N_{c}$
arguments do not explain the small width of exotic pentaquark
states.
\end{abstract}

\PACS{11.30.Rd, 12.39.Dc, 13.30.Eg, 14.20.--c}

\section{Introduction}
\label{intro}

There is still a lack of consensus whether the lightest member of
the exotic antidecuplet has been discovered \cite{exp}. Four
months after this conference results from high statistics G11
experiment at CLAS have been presented at the APS meeting with
negative result for the photoproduction of $\mathit{\Theta}^{+}$
on proton \cite{G11}. Even more problematic is the sighting of the
heaviest members of $\overline{10}$ that were seen only by NA49
experiment at CERN \cite{Xi}. These states were predicted within
the chiral soliton models
\cite{antidec,BieDotha,Mogil,DPP,Weigel}. Early
estimates of the antidecuplet octet splitting, $\mathit{\Delta}
M_{\overline{10}-8}\sim 600$~MeV, obtained in a specific
modification of the Skyrme model can be found in
Ref.~\cite{BieDotha}. The estimates of \emph{both}
$\mathit{\Theta}^{+}$ and $\mathit{\Xi}_{\overline{10}}$ masses
from the second order mass formulae obtained in the Skyrme model
in 1987 are in a surprising agreement with present experimental
findings \cite{Mogil}. Already at that time, however, the doubts
whether these predictions were trustworthy had been raised
\cite{BieDotha,Mogil,Klebanov}. Today they were scrutinized and
rephrased by other authors
\cite{cohenlargenc,pobylitsalargenc,bound}.

In 1997 the masses, as well as the decay widths of the exotic
states were estimated within the chiral quark soliton model
\cite{DPP}. One of the most striking predictions of this seminal
paper by Diakonov, Petrov and Polyakov \cite{DPP} was the small
width of antidecuplet states. Despite some misprints in this paper
(see \eg \cite{EKP,Arndt}) and the model dependent corrections,
the narrow width is one of the key features of the chiral model
predictions which is in line with recent experimental findings.

In this paper we examine the successes and problems of chiral
soliton models ability to predict properties of exotic baryons. In
Section \ref{sect:models} we argue that soliton models {\em are}
in fact quark models and explain the difference between
quark--soliton models and Skyrme model. Then  in
Sect.~\ref{sect:massses} we list different predictions for masses
of exotic baryons and discuss the $N_{c}$ counting for the mass
splittings. In Sect.~\ref{decays} we repeat the same analysis for
the decay widths. Summary is given in Sect.~\ref{sect:summary}.

\section{Soliton models} \label{sect:models}

Soliton models are often regarded as orthogonal to the quark
picture. Very often they are generally referred to as Skyrme type
models where only mesonic degrees are present. In this Section we
will demonstrate that they are deeply rooted in QCD, take into
account quark degrees of freedom maybe even in a more complete way
than the quark models themselves, and that they are fully
operative providing predictions of static baryon properties,
structure functions, skewed and off-forward amplitudes and
light-cone distribution amplitudes for baryons (for review see
{\em e.g.} Refs.~\cite{Diakonov,Weigel:1995cz,Christov:1995vm}),
not to mention
properties of pseudoscalar mesons \cite{mesons}. That of course
does not mean that they capture all physics, since --- for example
--- they do not posses confinement. They rely on large $N_{c}$
limit and chiral symmetry breaking. We shall also make distinction
between quark--soliton and Skyrme model.

Let us take as a starting point the chiral Lagrangian density
of the form%
\begin{equation}
\mathcal{L}=\overline{\psi}(i\,/\hspace{-6pt}\partial-M\,U^{\gamma_{5}}[\varphi
])\psi\label{Lagr}%
\end{equation}
which looks like a Dirac Lagrangian density for a massive fermion
$\psi$ if not for matrix $U$. In fact $\psi$ is a $3$-vector in
flavor space and also in color. Matrix
\[
U^{\gamma_{5}}=e^{({i}/{F_{\varphi})}\vec{\lambda}\cdot\vec{\varphi}%
\,\gamma_{5}}%
\]
parameterized by a set of pseudoscalar fields $\vec{\varphi}$ has
been introduced to restore chiral symmetry given by a global
multiplication of the
fermion field by a phase factor%
\begin{equation}
\psi\rightarrow e^{i\vec{\lambda}\cdot\vec{\alpha}\,\gamma_{5}}\psi
\,.\label{chitrans}%
\end{equation}
Indeed, the term $M\,\overline{\psi}\,\psi$ is not invariant under
(\ref{chitrans}), however
$M\,\overline{\psi}U^{\gamma_{5}}[\varphi]\psi$ is,
provided we also transform meson fields%
\begin{equation}
U^{\gamma_{5}}[\varphi]\rightarrow e^{-i\vec{\lambda}\cdot\vec{\alpha}%
\,\gamma_{5}}U^{\gamma_{5}}[\varphi]\,e^{-i\vec{\lambda}\cdot\vec{\alpha
}\,\gamma_{5}}.
\end{equation}
Since matrix $U$ \textquotedblleft lives\textquotedblright\ in
the flavor space, the color indices are here contracted producing
simply an overall factor $N_{c}$ in front of (\ref{Lagr}).

Lagrangian (\ref{Lagr}) does not contain kinetic term for meson
fields, so $\vec{\varphi}\,$'s are expressed in terms of fermion
fields themselves. The kinetic term appears only when we integrate
out the quark fields. Then the resulting effective action contains
only meson fields and can be organized in
terms of a derivative expansion%
\begin{eqnarray}
S_{\rm
eff}[\varphi]&=&\frac{F_{\varphi}^{2}}{4}\int\operatorname*{Tr}\left(
\partial_{\mu}U\,\partial^{\mu}U^{\dagger}\right) \nonumber\\ 
&&+\frac{1}{32e^{2}}%
\int\operatorname*{Tr}\left(  \left[  \partial_{\mu}U\,U^{\dagger}%
,\partial_{\nu}UU^{\dagger}\right]  ^{2}\right)
+\mathit{\Gamma}_{\rm WZ}+\ldots\,,
\label{SkS}%
\end{eqnarray}
where constants $F_{\varphi}$ and $e$ can be calculated from
(\ref{Lagr}) with an appropriate cut-off. $\mathit{\Gamma}_{\rm
WZ}$ is the Witten Wess--Zumino term which takes into account axial
anomaly and does not require regularization. Perhaps the most
important part are the ellipses which encode an infinite set of
terms that are effectively summed up by the fermionic model of
Eq.~(\ref{Lagr}). The truncated series of Eq.~(\ref{SkS}) is the
basis of the Skyrme model. Hence the Skyrme model is (a somewhat
arbitrary, because it does not include another possible 4
derivative term) approximation to (\ref{Lagr}).

At this point both models, chiral quark model of Eq.~(\ref{Lagr})
and Skyrme model of Eq.~(\ref{SkS}) (without the ``dots''), look like
mesonic theories devised to describe meson--meson scattering, for
example \cite{scattering}. Baryons are introduced in two steps,
following large $N_{c}$ strategy described by Witten in
Refs.~\cite{Witten}. First, one constructs a soliton solution,
\emph{i.e.} solution to the classical equations of motion that
corresponds to the extended configuration of the meson fields,
\emph{i.e.} to matrix $U$ which cannot be expanded in a power
series around unity. Second, since the classical soliton has no
quantum numbers (except baryon number, see below), one has to
quantize the system. Perhaps this quantization procedure, which
reduces both models to the nonrelativistic quantum system
analogous to the symmetric top \cite{ANW,antidec} with two moments
of inertia $I_{1,2}$, makes chiral-soliton models look odd and
counterintuitive.

It is not our purpose to give the full review of the soliton
models which can be found elsewhere
\cite{Diakonov,Weigel:1995cz,Christov:1995vm},
especially their connection with QCD, to some extent obvious from
Eq.~(\ref{Lagr}), was extensively reviewed in Ref.~\cite{Diakinst}.
Here we want to discuss the interconnection of the chiral soliton
models and quark models. In fact Lagrangian (\ref{Lagr}) \emph{is}
an interacting quark model, despite the fact that there are no
gluons there. The interaction proceeds through the chirally
invariant coupling $\overline{\psi}U^{\gamma_{5}}\psi$ and
information about gluons (which are integrated out) is encoded in
the coupling strength $M$ (constituent quark mass).

\begin{figure}[htb]
\begin{center}
\includegraphics*[scale=1.3]{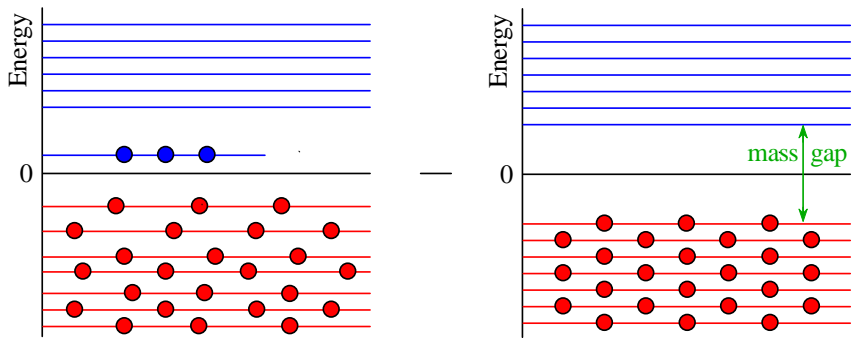}
\end{center}
\caption{Spectrum of the Dirac operator in the presence of the
valence level and without. The soliton energy is calculated as the
regularized difference of
two contributions}%
\label{fig:levels}%
\end{figure}

For the purpose of illustration it is convenient to use the
variational approach for the soliton solution
\cite{Diakonov:1988mg}. To this end one
uses a hedgehog ansatz for the static $U_{0}$ field:%
\begin{equation}
U_{0}=\left[
\begin{array}
[c]{cc}%
e^{i\vec{n}\cdot\vec{\tau}\,P(r)} & 0\\
0 & 1
\end{array}
\right], \label{hedgehog}%
\end{equation}
where $2\times2$ matrix in the upper left corner depends on one
variational function $P(r)=P(r/r_{0})$ characterized by an
effective size $r_{0}$ and being a subject to the boundary
conditions $P(0)=\pi$ and $P(\infty)=0$. For $r_{0}=0$ matrix
$U_{0}=1$ and the spectrum of the Dirac operator corresponding to
(\ref{Lagr}) looks like a spectrum of a free fermion of mass $M$
(see the right panel of Fig.~\ref{fig:levels}). Once we increase
the size $r_{0}$, the levels rearrange and one distinct level
\textquotedblleft sinks\textquotedblright\ rapidly into a mass
gap (see the left panel of Fig.~\ref{fig:levels}).

The energy of this \emph{interacting }fermionic system is given as
a sum of the valence level and the sea levels (filled levels of
the Dirac ``sea\textquotedblright) calculated
with respect to the vacuum ($r_{0}=0$) configuration as depicted
in Fig.~\ref{fig:levels}. Stable minimum is achieved for some
intermediate soliton size $r_{0}=r_{\rm sol}$, usually of the
order of a fermi, as an interplay between \emph{decreasing} energy
of the valence level and \emph{increasing} energy of the Dirac
sea. In this case the baryon number of the soliton is given simply
as the baryon number of the valence level.

An interesting limit can be considered by \emph{artificially}
tuning the size of the soliton to $r_{0}=0$ \cite{limit}. In this
limit the valence level goes back to the upper edge of the mass
gap and the contribution of the sea levels cancels out. Hence the
soliton in this limit looks like the constituent quark model.
Indeed, it has been shown that for $r_{0}\rightarrow0$ many static
observables calculated in the soliton model are in agreement with
the naive quark model predictions. These include $g_{A}=5/3,$
$\mathit{\Delta}\mathit{\Sigma}=1$ and $\mu_{n}/\mu _{p}=-2/3$.

 In the Skyrme model the soliton is constructed purely from the
mesonic field (\ref{hedgehog}). The baryon number is given as a
charge of the conserved topological current. Stabilization is
achieved by an interplay of the increasing energy of the kinetic
term and the decreasing energy of the Skyrme term (\ref{SkS}).
This reflects the main difference between quark--soliton and
Skyrme-soliton. Had we included all terms denoted by ellipses in
Eq.~(\ref{Lagr}) there would have been no minimum of energy as a
function of $r_{0}$.

The quantization on the other hand proceeds in both models almost
identically \cite{antidec}. The symmetric top Hamiltonian is
supplemented by a constraint coming from the $\mathit{\Gamma}_{\rm
WZ}$ term which selects SU(3) representations that contain states
of $Y=1$. Octet, decuplet, exotic antidecuplet and eikosiheptaplet
(\emph{i.e.} 27-plet) \cite{other27} are the lowest possible
representations satisfying this constraint. The difference between
the two models is buried in the analytical form of the
expressions for the symmetric top parameters (overall mass and
moments of inertia, \emph{etc.}). Some of them are identically
zero in the Skyrme model, whereas they are nonzero in the
quark--soliton model due to the valence level contribution.

Is the tower of representations satisfying constraint $Y=1$
infinite? Formally the answer is: yes, but physically: no, since
we have to revise assumptions which led us to the quantization of
the soliton. Namely, we have assumed \emph{rigid rotation} which
is (classically) very unlikely when the soliton angular velocities
become large. Two phenomena are expected: deformation of the
soliton and vibrations. Deformation will lead to instabilities
resulting in radiation of pions (Goldstone bosons in general).
Fast rotating solitons will
have a cigar-like shape and will lie on linear Regge trajectories
\cite{Diakonov:2003uv}.

As discussed above there is only one representation of given
dimension in the allowed series of representations selected by the
Wess--Zumino constraint. So there is only one (nonexotic) octet,
while the quark models inevitably require a cryptoexotic octet
together with antidecuplet. Of course, the octet
is not missing; it has to be of different origin. So far we have
constructed only rotational states, however, there will be also
vibrations.

Explicit construction of the vibrational states in the Skyrme
model (with the dilaton field) has been carried on by Weigel in
Refs.~\cite{Weigel}. In this approach only one mode, namely the
``breathing\textquotedblright\ mode of the
soliton was quantized and a subsequent mixing with other states
was investigated.

\section{Mass estimates} \label{sect:massses}

In order to estimate the mass of $\mathit{\Theta}^{+}$ we have to
know two quantities, namely the strange moment of inertia, $I_{2}$
which contributes to antidecuplet--octet splitting
\begin{equation}
\mathit{\Delta} M_{\overline{10}-8}=\frac{3}{2I_{2}}\label{10bar8}%
\end{equation}
and the energy shift due to the nonzero strange quark mass. It has
been observed by Guadagnini \cite{Guad} that the mass splittings
in the Skyrme model are well
described in terms of 2 parameter effective Hamiltonian%
\[
H^{^{\prime}}=\alpha D_{88}^{(8)}+\beta Y\,,
\]
where $\alpha\sim\mathit{\Sigma}_{\pi N}$. However $\beta\equiv0$
in the minimal Skyrme model and the spectrum cannot be well
described in the first order perturbation in $H^{\prime}$. Second
order
correction, which can be schematically written as \cite{Mogil}%
\begin{equation}
\mathit{\Delta} E_{2}\sim-I_{2}\times\alpha^{2}%
\end{equation}
mimics the nonzero $\beta$. This second order correction cannot be
smaller than the typical mismatch of the first order
$\mathit{\Delta} M$. Therefore, there is a lower bound on
$\mathit{\Delta} E_{2}$, which translates into an upper bound on
$\mathit{\Delta} M_{\overline{10}-8}$. The antidecuplet cannot be
too heavy. On the other hand $\mathit{\Delta} E_{2}$ cannot be too
large for consistency reasons, hence too light antidecuplet is
also excluded. This is how the original prediction
$M_{\mathit{\Theta}^{+}}\sim1535$ MeV was obtained \cite{Mogil}.
The updated results of this analysis are given in Table~I.
\begin{table}[ptb]
{\small
\rightline{TABLE I}

\vspace{0.3cm}\noindent
Masses of baryons obtained by minimizing the square
deviation with respect to $M_{\rm sol},\;\; I_1\;\; I_2$ for fixed
$\alpha$.

\vspace{0.2cm}
\begin{center}%
\begin{tabular}
[c]{lccr}
\hline\hline
\w& exp. & $\alpha=720$ MeV & push\\[1mm]
\hline
\w$N$ & \ 939 & \ 915 & $-23$\\
$\mathit{\Lambda}$ & 1116 & 1090 & $-26$\\
$\mathit{\Sigma}$ & 1193 & 1214 & $+21$\\
$\mathit{\Xi}$ & 1318 & 1323 & $+5$\\[1mm]
\hline
\w$\mathit{\Delta}$ & 1232 & 1231 & $-1$\\
$\mathit{\Sigma}^{\ast}$ & 1385 & 1389 & $+4$\\
$\mathit{\Xi}^{\ast}$ & 1533 & 1535 & $+2$\\
$\mathit{\Omega}^{-}$ & 1672 & 1662 & $-10$\\[1mm]
\hline
\w$\mathit{\Theta}^{+}$ & 1540 & 1535 & $-5$\\
N$_{\overline{10}}^{\ast}$ &  & 1667 & \\
$\mathit{\Sigma}_{\overline{10}}^{\ast}$ &  & 1751 & \\
$\mathit{\Xi}_{\overline{10}}^{\ast}$ & 1862 & 1784 &
$-78$\\
\end{tabular}
\end{center}
}
\label{tabI}%
\end{table}

In the quark--soliton models one chooses different path. Instead of
going to the second order in perturbative expansion in $m_{s}$ one
calculates nonleading terms in $1/N_{c}$ \cite{Blotzsu3}. This
generates $\beta\neq0$ from the beginning and the lower bound on
$I_{2}$ does not exist. One can try either to constrain $I_{2}$ by
identifying some known nucleon resonance with
$N_{\overline{10}}^{\ast}$ , as it was done in Ref.~\cite{DPP}, or
resort to explicit model calculations which, however, cover rather
broad range of allowed values \cite{EKP}.

In the original paper of Diakonov, Petrov and Polyakov \cite{DPP}
the value of $I_{2}$ was fixed by the identification of
$N_{\overline{10}}^{\ast}$ with $N^{\ast }(1710)$ and the equal
spacing in antidecuplet by adopting the value of $45$ MeV for
$\mathit{\Sigma}_{\pi N}$.
\begin{figure}[htb]
\begin{center}
\includegraphics*[scale=2]{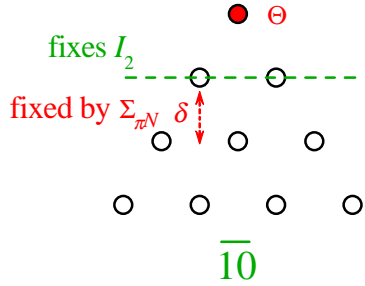}
\end{center}
\caption{Antidecuplet of SU(3) flavor including
$\mathit{\Theta}^{+}$. In Ref.~\cite{DPP} $I_{2}$ was fixed by
$N^{\ast}(1710)$ and the splitting
$\delta$ by fixing $\mathit{\Sigma}_{\pi N}$. }%
\label{fig:a10}%
\end{figure}

Today we would take a different approach. We would use
$\mathit{\Theta}^{+}$ rather than $N_{\overline{10}}^{\ast}$ to
fix the average $\overline{10}$ mass. In a recent paper \cite{EKP}
it has been shown that the set of parameters of the
symmetry breaking Hamiltonian%
\begin{equation}
\hat{H}^{\prime}=\alpha D_{88}^{(8)}+\beta Y+\frac{\gamma}{\sqrt{3}}%
D_{8i}^{(8)}\hat{S}_{i}\label{Hprim}%
\end{equation}

\noindent(where $D_{88}^{(8)}$ are SU(3) Wigner matrices, $Y$ is
hypercharge and $\hat{S}_{i}$ is the collective spin operator
\cite{Blotzsu3}) which reproduces well the nonexotic spectra, as
well as the measured mass of the $\mathit{\Theta}^{+}(1540)$, can
be parametrized as follows\footnote{We use
here $m_{s}/(m_{u}+m_{d})=12.9$ \cite{Leutwyler}.}:%
\begin{equation}
\alpha=336.4-12.9\,\mathit{\Sigma}_{\pi N}\,,\quad\beta
=-336.4+4.3\,\mathit{\Sigma}_{\pi N},\quad\gamma=-475.94+8.6\,\mathit{\Sigma
}_{\pi N}\,.\label{albega}%
\end{equation}
Moreover, the inertia parameters which describe the representation splittings%
\begin{equation}
\mathit{\Delta}M_{{10}-8}=\frac{3}{2I_{1}}\,,\quad \mathit{\Delta}M_{{\overline
{10}}-8}=\frac{3}{2I_{2}}%
\end{equation}
take the following values (in MeV)%
\begin{equation}
\frac{1}{I_{2}}=152.4\,,\quad\frac{1}{I_{2}}=608.7-2.9\,\mathit{\Sigma}_{\pi
N}\,.
\end{equation}
If, furthermore, one imposes additional constraint that
$M_{\mathit{\Xi }_{\overline{10}}}=1860$~MeV, then
$\mathit{\Sigma}_{\pi N}=73$ MeV \cite{EKP} (see also \cite{Schw})
in agreement with recent experimental estimates \cite{sigma}.

Hamiltonian (\ref{Hprim}) introduces mixing between different
representations\break\cite{EKP,oper}: \vspace{-0.3cm}
\begin{align}
\left\vert B_{8}\right\rangle  &  =\left\vert 8_{1/2},B\right\rangle
+c_{\overline{10}}^{B}\left\vert \overline{10}_{1/2},B\right\rangle
+c_{27}^{B}\left\vert 27_{1/2},B\right\rangle \,,\nonumber\\
\left\vert B_{10}\right\rangle  &  =\left\vert 10_{3/2},B\right\rangle
+a_{27}^{B}\left\vert 27_{3/2},B\right\rangle +a_{35}^{B}\left\vert
35_{3/2},B\right\rangle \,,\nonumber\\
\left\vert B_{\overline{10}}\right\rangle  &  =\left\vert \overline{10}%
_{1/2},B\right\rangle +d_{8}^{B}\left\vert 8_{1/2},B\right\rangle +d_{27}%
^{B}\left\vert 27_{1/2},B\right\rangle +d_{\overline{35}}^{B}\left\vert
\overline{35}_{1/2},B\right\rangle \,,\label{admix}%
\end{align}
where $\left\vert B_{\mathcal{R}}\right\rangle $ denotes the state which
reduces to the SU(3) representation $\mathcal{R}$ in the formal limit
$m_{s}\rightarrow0$. The $m_{s}$ dependent (through the linear $m_{s}$
dependence of $\alpha$, $\beta$ and $\gamma$) coefficients $c_{\mathcal{R}%
}^{B}$, $d_{\mathcal{R}}^{B}$ and $a_{\mathcal{R}}^{B}$ in
Eq.~(\ref{admix}) can be found \emph{e.g.} in Ref.~\cite{EKP}.

Although the model seems to describe the spectrum of
antidecuplet rather well (assuming that $\mathit{\Theta}^{+}$ and
$\mathit{\Xi}_{\overline{10}}$ exist and have masses as discussed
in the Introduction), we encounter here the first potential
problem. Namely, the exotic--nonexotic mass splitting
(\ref{10bar8})
reads in fact%
\begin{equation}
\mathit{\Delta}M_{\overline{10}-8}=\frac{N_{c}}{2I_{2}}\sim\mathcal{O}%
(1)\label{10bar8Nc}%
\end{equation}
whereas%
\begin{equation}
\mathit{\Delta}M_{{10}-8}=\frac{3}{2I_{1}}\sim\mathcal{O}(N_{c}^{-1}%
)\,.\label{108Nc}%
\end{equation}
This $N_{c}$ counting is in fact born by experiment%
\begin{equation}
\mathit{\Delta}M_{{10}-8}\simeq230\;\text{MeV}\,,\qquad\mathit{\Delta
}M_{\overline{10}-8}\simeq600\;\text{MeV\,,}%
\end{equation}
however, it poses a serious problem to the validity of the
quantization procedure for exotic states in large $N_{c}$ limit.
Indeed, in the imaginary world of very large $N_{c}$ all nonexotic
states are degenerate, whereas exotic ones are split by a quantity
of the order $\mathcal{O}(1)$, similarly to the vibrations with
which they will mix. This fact although known already in the late
80's have been recently revised critically in the literature
\cite{cohenlargenc,pobylitsalargenc}

\section{Decay widths} \label{decays}

In the decay of $\mathit{\Theta}^{+} \rightarrow NK$ the kaon
momentum in the rest
frame of the decaying particle%
\begin{equation}
p_{K}=267\ \text{MeV}\label{pK}%
\end{equation}
is almost identical to the pion momentum in $\mathit{\Delta}$ decay%
\begin{equation}
p_{\pi}=225\ \text{MeV.}\label{ppi}%
\end{equation}
One would, therefore, naively expect that the decay width of
$\mathit{\Theta}^{+}$ should be at least as large as the one of
$\mathit{\Delta}$ or even larger, since no suppression coming from
the overlap of the wave functions is expected. Indeed, pentaquark
states have --- naively --- so called fall-apart modes. One of the
chief predictions of the quark--soliton models is that the
$\mathit{\Theta}^{+} $ width, contrary to the naive expectations,
is very small \cite{DPP}. This prediction stimulated experimental
searches.

Whilst the mass spectra discussed in the previous section are given as
systematic expansions in both $N_{c}$ and $m_{s}$ in a theoretically
controllable way, reliable predictions for the decay widths cannot be
organized in a similar manner. In fact the decay width is calculated by means
of the formula for the decay width for $B\rightarrow B^{\prime}+\varphi$:%
\begin{equation}
\mathit{\Gamma}_{B\rightarrow B^{\prime}+\varphi}=\frac{1}{8\pi}%
\frac{p_{\varphi}}{M\,M^{\prime}}\overline{\mathcal{M}^{2}}=\frac{1}{8\pi
}\frac{p_{\varphi}^{3}}{M\,M^{\prime}}\overline{\mathcal{A}^{2}}%
\label{Gammadef}%
\end{equation}
up to linear order in $m_{s}$. The \textquotedblleft bar\textquotedblright%
\ over the amplitude squared denotes averaging over initial and
summing over final spin (and, if explicitly indicated, over
isospin). Anticipating linear momentum dependence of the decay
amplitude $\mathcal{M}$ we have introduced reduced amplitude
$\mathcal{A}$ which does not depend on the kinematics, \ie on the
meson momentum $p_{\varphi}$. For the discussion of the validity
of (\ref{Gammadef}) see \cite{EKP}.

Soliton models can be used to calculate the matrix element $\mathcal{M}$. In
order to match former normalization \cite{DPP} we shall define the decay
amplitude as%
\begin{align}
\mathcal{M}_{B\rightarrow B^{\prime}+\varphi}  & =\left\langle B^{\prime
}\right\vert \hat{O}_{\varphi}^{(8)}\left\vert B\right\rangle \nonumber\\
& =3\left\langle B^{\prime}\right\vert G_{0}D_{\varphi i}-G_{1}d_{ibc}%
\,D_{\varphi b}^{(8)}\hat{S}_{c}-\frac{G_{2}}{\sqrt{3}}D_{\varphi8}^{(8)}%
\hat{S}_{i}\left\vert B\right\rangle \times p_{\varphi}^{i}%
\,,\label{Op}%
\end{align}
where the sum over the repeated indices is assumed: $i=1,2,3$ and
$b,c=4,\ldots7 $. It is assumed that coupling constants
$G_{0,1,2}$ can be related to the elements of the axial current by
means of the generalized Goldberger--Treiman
relations \cite{DPP}. Explicitly%
\[
\mathit{\Gamma}_{B\rightarrow B^{\prime}+\varphi}=\frac{3\, G_{\mathcal{R}}^{2}%
}{8\pi M\,_{B}M_{B^{\prime}}}C_{B\rightarrow B^{\prime}+\varphi}^{\mathcal{R}%
}\,p_{\varphi}^{3}\,.
\]
For antidecuplet decays ($\mathcal{R}=\overline{10}$):%
\begin{equation}
G_{\overline{10}}\;=G_{0}-G_{1}-{\frac{1}{2}}G_{2}\,,\quad
C_{\mathit{\Theta} ^{+}\rightarrow
N+K}^{\overline{10}}=\frac{1}{5}\,,
\end{equation}
whereas for decuplet ($\mathcal{R}=10$):%
\begin{equation}
G_{10}\;=G_{0}+{\frac{1}{2}}G_{2},\quad C_{\mathit{\Delta}\rightarrow N+\pi}^{10}%
=\frac{1}{5}\,.
\end{equation}
Hence the suppression of antidecuplet decay width may come only from the
cancellation between $G_{0,1,2}$ entering $G_{\overline{10}}$. Indeed, in the
nonrelativistic small soliton limit discussed above one can show that
$G_{1}/G_{0}=4/5$, $G_{2}/G_{0}=2/5$ and $G_{\overline{10}}\equiv0$! Although
nonintuitive this cancellation explains the small width of antidecuplet as
compared to the one of $10$ for example.

One problem concerning this cancellation is that formally%
\begin{equation}
G_{0}\sim\mathcal{O}(N_{c}^{3/2})+\mathcal{O}(N_{c}^{1/2}),\quad G_{1,2}%
\sim\mathcal{O}(N_{c}^{1/2})
\end{equation}
and it looks as if the cancellation were accidental as it occurs
between terms of different order in $N_{c}$. That this is not the
case was shown in Ref.~\cite{MPGamma}. Indeed for arbitrary $N_{c}$
antidecuplet $\overline{10}=(0,3)$ generalizes to
$``\overline{10}"=(0,\frac{N_{c}+3}{2})$, decuplet
$``10"=(3,\frac{N_{c}-3}{2})$ and octet $``8"=(1,\frac{N_{c}-1}{2})$
\cite{largereps}, and the pertinent Clebsch--Gordan
coefficients in fact depend on $N_{c}$:%
\begin{equation}
G_{``\overline{10}"}=G_{0}-\frac{N_{c}+1}{4}G_{1}-\frac{1}{2}G_{2}\,.
\end{equation}
So the subleading $G_{1}$-term is enhanced by additional factor
of $N_{c}$ and the cancellation is consistent with $N_{c}$ counting.
Moreover%
\begin{align}
C_{\mathit{\Theta}^{+}\rightarrow N+K}^{``\overline{10}"}  & =\frac{3(N_{c}+1)}%
{(N_{c}+3)(N_{c}+7)}\sim\mathcal{O}\left(\frac{1}{N_{c}}\right)\,,\nonumber\\
C_{\mathit{\Delta}\rightarrow N+\pi}^{``10"}  & =\frac{(N_{c}-1)(N_{c}+5)}%
{2(N_{c}+1)(N_{c}+7)}\sim\mathcal{O}(1)
\end{align}
and it looks like the antidecuplet width were suppressed with respect to
decuplet. Unfortunately the phase space factor $p_{\varphi}^{3}$ spoils this
counting. Indeed, because of (\ref{10bar8Nc}) and (\ref{108Nc})%
\begin{equation}
p_{\pi}\sim\mathcal{O}\left(\frac{1}{N_{c}}\right)\,,\quad p_{K}\sim\mathcal{O}%
(1)\label{pscaling}%
\end{equation}
and consequently%
\begin{equation}
\mathit{\Gamma}_{\mathit{\Delta}\rightarrow N+\pi}\sim\mathcal{O}\left(\frac{1}{N_{c}^{2}}%
\right),\quad\mathit{\Gamma}_{\mathit{\Theta}^{+}\rightarrow
N+K}\sim\mathcal{O}(1)
\end{equation}
in the chiral limit. This $N_{c}$ counting contradicts
experimental findings which suggest
$\mathit{\Gamma}_{\mathit{\Theta}^{+}\rightarrow
N+K}\ll\mathit{\Gamma}_{\mathit{\Delta}\rightarrow N+\pi}$.

\begin{figure}[htb]
\begin{center}
\includegraphics*[scale=0.8]{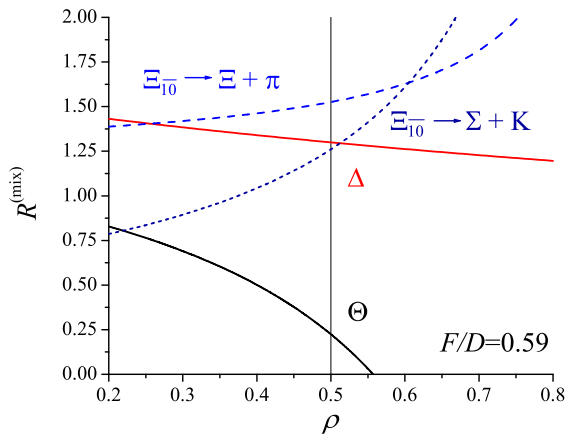}
\end{center}
\caption{ The correction factors $R^{\rm (mix)}$ due to SU(3)-breaking
representation mixing for the decays discussed in the text, as functions of
the parameter $\rho\equiv G_{1} / G_{0}$. }%
\label{fig:rmix}%
\end{figure}
A few comments are in order. Firstly, let us note that in Nature
neither $\pi$ nor $K$ mesons are massless and both $p_{\pi}$ and
$p_{K}$ are of the same order of 230 MeV (\ref{pK}), (\ref{ppi}) and
the scaling (\ref{pscaling}) does not hold. So for $m_{\pi}\neq0$
and $m_{K}\neq0$ both meson momenta scale as $N_{c}^{0}$, however
$\mathit{\Delta}$ does not decay, because in the large $N_{c}$
limit it is degenerate with nucleon and cannot emit a massive
particle, whereas $\mathit{\Theta}^{+}$ does decay. It is an
instructive example how subtle is an
interplay of theoretical limits $N_{c}\rightarrow\infty$ and $m_{q}%
\rightarrow0$. Secondly, (\ref{pscaling}) holds only if the cancellation
$G_{\overline{10}}=0$ is not exact. Let us suppose that the leading $N_{c}$
power cancels in such a way that $G_{\overline{10}}\sim\mathcal{O}(N_{c}%
^{1/2})$ rather than $\mathcal{O}(N_{c}^{3/2})$. That would make
$\mathit{\Gamma}_{\mathit{\Theta}^{+}\rightarrow
N+K}\sim\mathcal{O}(1/N_{c}^{2})$, \emph{i.e.} of the same order
as $\mathit{\Gamma}_{\mathit{\Delta}\rightarrow N+\pi}$. Finally
let us remark that there is further suppression of
$\mathit{\Gamma}_{\mathit{\Theta}^{+}\rightarrow N+K}$ coming from
the mixing (\ref{admix}). This is illustrated in
Fig.~\ref{fig:rmix} where we plot multiplicative correction factor
$R^{\rm (mix)}$ as a function of the parameter $\rho\equiv G_{1} /
G_{0}$. Phenomenological value of $\rho\sim0.5$ \cite{EKP}.

\section{Summary} \label{sect:summary}

The solitonic approach to baryons is very successful, as it
describes not only spectra and other static properties, but also
structure functions, skewed and nonforward parton distributions
and also ligth-cone distribution amplitudes. However, it relies on
many approximations. Firstly, it is based on an ansatz and as such
one must check the self-consistency of the approach. Evidently,
following arguments by Witten \cite{Witten}, solitonic solutions
are justified only for large $N_{c}$. Indeed, looking \eg
at Eq.~(\ref{Lagr}) a calculation of baryonic properties requires
functional integration over both bosonic ($\varphi$) and fermionic
($\psi$) degrees of freedom. In order to apply the stationary
phase approximation, as it is commonly done in the soliton models,
one has to omit the bosonic functional integral, using instead the
\emph{background} bosonic field that minimizes the effective
action of the system. This is only justified for large $N_{c}$,
where the bosonic fluctuations are suppressed. Secondly, soliton
quantization proceeds by quantizing the rotations in space and
flavor space. To this end one assumes the rotational motion to be
adiabatic. This means that angular velocities go like
$J/{I}\sim1/N_{c}$ yielding frequencies (and hence excitation
energies) of order $\mathcal{O}(N_{c}^{-1})$. This, in turn,
implies a Born--Oppenheimer separation of the slow collective
rotational motion from the faster modes associated with
vibrations. Because of this scale separation the collective
rotational modes can be quantized separately from the intrinsic
vibrations. While this procedure has been applied with great
success to many properties of the nonexotic baryons it has been
criticized as far as exotic multiplets are concerned.

The question \cite{cohenlargenc,pobylitsalargenc} here is whether
the rigid-rotor type semiclassical projection can be applied to
exotic states. The fact that the standard semiclassical
quantization gave excitation energies of the order
$\mathcal{O}(N_{c}^{0})$ for exotic states (\ref{10bar8Nc}) means
that the approach is not justified for such states unless further
arguments can be invoked. In contrast, in view of
Eq.~(\ref{108Nc}), rigid-rotor quantization is certainly justified
for the non-exotic states.

Diakonov and Petrov \cite{diakpetlargenc} have argued that while
it is true that at large $N_{c}$ rotational excitations are
comparable to vibrational or radial excitations of baryons, both
non-exotic and exotic, the corrections due to the coupling between
rotations and vibrations die out as $1/N_{c}$. The collective
rotational quantization description fails only when the
exoticness, \emph{i.e.} the number of $q\bar{q}$ valence pairs
needed to construct the quantum numbers of a given state, becomes
comparable to $N_{c}$. The newly discovered $\mathit{\Theta}^{+}$
baryon belongs to the antidecuplet of exoticness $=1$. The larger
$N_{c}$, the more accurate would be its description as a
rotational state of a chiral soliton. Diakonov and Petrov
\cite{diakpetlargenc} support their estimates by considering a
simple model consisting of a charged particle in the field of a
monopole. However, if one generalizes the model by considering two
charged particles interacting by a harmonic potential and moving
in the field of a monopole, the coupling of rotational and
vibrational degrees of freedom of this system is by no means
vanishing but strong \cite{pobylitsalargenc}.

As we have discussed in Sect.~\ref{decays} large-$N_{c}$ arguments
apply also to the width of the baryons considered, because if the
approach is justified, then at a formal level the width must
approach zero at large $N_{c}$. Of course, for non-exotic states
such as the decuplet, this is true. The reason is simply phase
space (\ref{pscaling}). Unfortunately, as shown recently in
Ref.~\cite{MPGamma} the width of the $\mathit{\Theta}^{+}$ as
calculated via the standard collective rotational approach is of
the order $N_{c}^{0}$ in the chiral limit. This demonstrates that
the procedure is not self consistent on the basis of pure
large-$N_{c}$ arguments. Thus, if the width of the
$\mathit{\Theta}^{+}$ is really small, as the experiments
indicate, there must be particular dynamical reasons for the
smallness, which exist on top of what is required for the validity
of the large-$N_{c}$ expansion alone. In this context the
cancellation leading to $G_{\overline{10}}=0$ in the small soliton
limit is of particular importance.

For completeness one should also mention another approach to the
quantization of chiral solitons based on the assumption that SU(3)
symmetry is strongly broken \cite{bound}. This approach, known as
a bound-state approach, was recently applied to
$\mathit{\Theta}^{+}$ by Klebanov {\em et al}. \cite{Klebanov}.
These authors reconsider the relationship between the SU(3)
rigid-rotator and the bound-state approach to strangeness in the
chiral soliton models. For non-exotic ${S}=-1$ baryons the
bound-state approach matches for small kaon mass $m_{K}$ onto the
rigid-rotator approach, and the bound-state mode turns into the
rotator zero-mode. However, for small $m_{K}$, there are no
${S}=+1$ kaon bound states or resonances in the spectrum
(unless $m_{K}\equiv0$). This shows that for large $N_{c}$ and
small (but non-zero) $m_{K}$ the exotic state is an artifact of
the rigid-rotator approach. An ${S}=+1$ near-threshold
state with the quantum numbers of the $\mathit{\Theta}^{+}$
pentaquark comes into existence only when SU(3) breaking is
sufficiently strong or vector mesons are introduced
\cite{Park:2004yf}. Therefore, one argues that pentaquarks are not
generic predictions of the chiral soliton models.

\bigskip
The present work is supported by the Polish State Committee for Scientific
Research (KBN) under grant 2 P03B 043 24.

\end{document}